\documentclass[10pt,conference]{IEEEtran}
\usepackage{cite}
\usepackage{subfigure}
\usepackage{graphicx}
\usepackage{mathrsfs}
\usepackage{amsmath}
\usepackage{amssymb}
\usepackage{algorithm2e}
\usepackage{paralist}

\makeatletter
\def\ps@headings{%
\def\@oddhead{\mbox{}\scriptsize\rightmark \hfil \thepage}%
\def\@evenhead{\scriptsize\thepage \hfil \leftmark\mbox{}}%
\def\@oddfoot{}%
\def\@evenfoot{}}
\makeatother 

\newtheorem{theorem}{{Theorem}}
\newtheorem{lemma}[theorem]{{Lemma}}

\newtheorem{definition}{{Definition}}

\allowdisplaybreaks

\newcommand{\nix}[1]{}

\begin{document}
\title{Raptor Codes Based Distributed Storage Algorithms for  Wireless Sensor Networks}
\author{\authorblockN{Salah A. Aly}
\authorblockA{
Department of Computer Science\\
Texas A\&M University\\
College Station, TX 77843, USA \\
Email: salah@cs.tamu.edu} \and
\authorblockN{Zhenning Kong}
\authorblockA{Department of Electrical Engineering\\
Yale University\\
New Haven, CT 06520, USA \\
Email: zhenning.kong@yale.edu} \and
\authorblockN{Emina Soljanin}
\authorblockA{Bell Laboratories\\
Alcatel-Lucent\\
Murray Hill, NJ 07974, USA \\
Email: emina@lucent.com}}
\maketitle

\begin{abstract}
We consider a distributed storage problem in a large-scale wireless
sensor network with $n$ nodes among which $k$ acquire
(sense) independent data. The goal is to disseminate the
acquired information throughout the network so that each of the $n$
sensors stores one possibly coded packet and the original $k$ data
packets can be recovered later in a computationally simple way from
any $(1+\epsilon)k$ of nodes for some small $\epsilon>0$. We propose
two Raptor codes based distributed storage algorithms for solving
this problem. In the first algorithm, all the sensors have the
knowledge of $n$ and $k$. In the second one, we assume that no
sensor has such global information.
\end{abstract}

\section{Introduction}\label{sec:into}

We consider a distributed storage problem in a large-scale wireless
sensor network with $n$ nodes among which $k$ sensor nodes acquire
(sense) independent data. Since sensors are usually vulnerable due to limited
energy and hostile environment, it is desirable to disseminate the
acquired information throughout the network so that each of the $n$
sensors stores one possibly coded packet and the original $k$ source
packets can be recovered later in a computationally simple way from
any $(1+\epsilon)k$ of nodes for some small $\epsilon>0$. No
sensor knows locations of any other sensors except for
their own neighbors, and they do not maintain any routing
information (e.g., routing tables or network topology).

Algorithms that solve such problems using coding in a centralized
way are well known and understood.
In a sensor network, however, this
is much more difficult, since we need to find a strategy to
distribute the information from multiple sources throughout the
network so that each sensor admits desired statistics of data.
In~\cite{lin07a}, Lin~\emph{et al.} proposed an algorithm that uses
random walks with traps to disseminate the source packets in a
wireless sensor network. To achieve desired code degree
distribution, they employed the Metropolis algorithm to specify
transition probabilities of the random walks. While the proposed
methods in~\cite{lin07a} are promising, the knowledge of the total
number of sensors $n$ and sources $k$ are required. Another type of
global information, the maximum node degree (i.e., the maximum
number of neighbors) of the graph, is also required to perform the
Metropolis algorithm. Nevertheless, for a large-scale sensor
network, these types of global information may not be easy to obtain
by each individual sensor, especially when there is a possibility of
change of topology. 

In~\cite{aly08a,aly08b}, we proposed Luby Transform (LT) codes based
distributed storage algorithms for large-scale wireless sensor
networks to overcome these difficulties. In this paper, we extend
this work to Raptor codes and demonstrate their performance. Particularly, we propose two new decentralized algorithms, Raptor Code Distributed Storage (RCDS-I) and
(RCDS-II), that distribute information sensed by k source nodes
to n nodes for storage based on Raptor codes. In RCDS-I, each node has limited
global information; while in RCDS-II, no global information is
required. We compute the computational encoding and decoding
complexity of these algorithms as well as evaluate their performance by simulation.

\section{Wireless Sensor Networks and Fountain Codes}\label{sec:model_Rcodes}

\subsection{Network Model}

Suppose that the wireless sensor network consists of $n$ nodes that
are uniformly distributed at random in a region
$\mathcal{A}=[L,L]^2$. Among these $n$ nodes, there are $k$ source
nodes that have information to be disseminated throughout the
network for storage. These $k$ nodes are uniformly and independently
chosen at random among the $n$ nodes. Usually, the fraction of
source nodes. 

We assume that no node has knowledge about the locations of
other nodes and no routing table is maintained, and thus that the
algorithm proposed in~\cite{dimakis05} cannot be applied. Moreover,
besides the neighbor nodes, we assume that each node has limited or
no knowledge of global information. The limited global information
refers to the total number of nodes $n$, and the total number of
sources $k$. Any further global information, for example, the
maximal number of neighbors in the network, is not available. Hence,
the algorithms proposed in~\cite{lin07a,lin07b,kamra06} are not applicable.

\begin{definition}(Node Degree)
Consider a graph $G=(V,E)$, where $V$ and $E$ denote the set of
nodes and links, respectively. Given $u,v\in V$, we say $u$ and $v$
are \emph{adjacent} (or $u$ is adjacent to $v$, and vice versa) if
there exists a link between $u$ and $v$, i.e., $(u,v)\in E$. In this
case, we also say that $u$ and $v$ are \emph{neighbors}. Denote by
$\mathcal{N}(u)$ the set of neighbors of a node $u$. The number of
neighbors of a node $u$ is called the \emph{node degree} of $u$, and
denoted by $d_n(u)$, i.e., $|\mathcal{N}(u)|=d_n(u)$. The \emph{mean
degree} of a graph $G$ is then given by $\mu =
\frac{1}{|V|}\sum_{u\in G}d_n(u)$.
\end{definition}

\subsection{Fountain Codes and Raptor Codes}

\begin{definition}(Code Degree)
For Fountain codes, the number of source blocks used to generate an
encoded output $y$ is called the code degree of $y$, and denoted by
$d_c(y)$. The code degree distribution $\Omega(d)$ is the
probability distribution of $d_c(y)$.
\end{definition}

\begin{figure}[t!]
\centering
\includegraphics[scale=0.35]{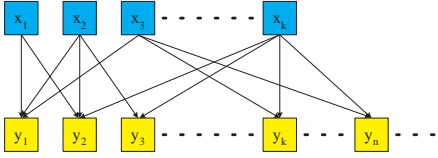}
\caption{The encoding operations of Fountain codes: each output is
obtained by XORing $d$ source blocks chosen uniformly and
independently at random from $k$ source inputs, where $d$ is drawn
according to a probability distribution $\Omega(d)$.}
\label{fig:fountaincodes} \vspace{-.2in}
\end{figure}

For $k$ source blocks $\{x_1,x_2,\ldots,x_k\}$ and a probability
distribution $\Omega(d)$ with $1 \leq d \leq k$, a Fountain code
with parameters $(k,\Omega)$ is a potentially limitless stream of
output blocks $\{y_1,y_2,...\}$. Each output block is obtained by
XORing $d$ randomly and independently chosen source blocks, where
$d$ is drawn from a degree distribution $\Omega(d)$. This is
illustrated in Fig.~\ref{fig:fountaincodes}.

\begin{figure}[t!]
\centering
\includegraphics[scale=0.35]{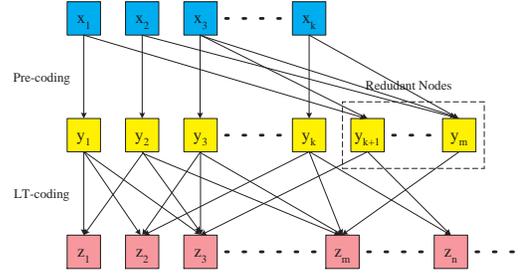}
\caption{The encoding operations of raptor codes: $k$ source blocks
are first encoded to $m$ pre-coding output blocks by LDPC coding,
and then the final encoded output blocks are obtained by applying LT
codes with these $m$ pre-coding output blocks with degree
distribution $\Omega_{r}(d)$.}\label{fig:raptorcodes}\vspace{-.2in}
\end{figure}

Raptor codes are a class of Fountain codes with linear encoding and
decoding complexity~\cite{shokrallahi04,shokrallahi06}. The key idea
of Raptor codes is to relax the condition that all input blocks need
to be recovered. If an LT code needs to recover only a constant
fraction of its input blocks, its  decoding complexity is $O(k)$,
i.e., linear time decoding. Then, we can recover all input blocks by
concatenating a traditional erasure correcting code with an LT code.
This is called pre-coding in Raptor codes, and can be accomplished
by a modern block code such as  LDPC codes. This process is
illustrated in Fig.~\ref{fig:raptorcodes}.

The pre-code $\mathcal{C}_m$  used in this paper is the randomized
LDPC (Low-Density Parity-Check) code that is studied as one type of
pre-code in~\cite{shokrallahi06}. In this randomized LDPC code, we
have $k$ source blocks and $m$ pre-coding output blocks. Each source
block chooses $d$ pre-coding output blocks uniformly independently
at random, where $d$ is drawn from a distribution $\Omega_{L}(d)$.
Each pre-coding output blocks combines the ``incoming'' source
blocks and obtain the encoded output.

The code degree distribution $\Omega_r(i)$ of Raptor codes for LT
coding is a modification of the Ideal Soliton distribution and given
by
\begin{equation}\label{eq:Raptor-distribution}
\Omega_r(i)=\left\{ \begin{array}{ll} \displaystyle \frac{\rho}{1+\rho}, & i=1,\\
\displaystyle \frac{1}{i(i-1)(1+\rho)}, & i=2,...,D, \\
\displaystyle\frac{1}{D(1+\rho)}, & i=D+1,\end{array}\right.
\end{equation}
where $D=\lceil 4(1+\epsilon)\epsilon \rceil$ and $\rho =
(\epsilon/2)+(\epsilon/2)^2$.

The following result provides the performance of the Raptor
codes~\cite{shokrallahi04,shokrallahi06}.

\begin{lemma}[{Shokrollahi~\cite{shokrallahi04,shokrallahi06}}]\label{Lemma:Decoding-Raptor-Codes}
Let $R_0 = (1+\epsilon/2)/(1 +\epsilon)$, and $\mathcal{C}_m$ be the
family of codes of rate $R_0$. Then, the Raptor code with pre-code
$\mathcal{C}_m$ and LT codes with degree distribution $\Omega_r(d)$
has a linear time encoding algorithm. With $(1+\epsilon)k$ encoded
output blocks, the BP decoding algorithm has a linear time
complexity. More precisely, the average number of operations to
produce an output symbol is $O(\log(1/\epsilon))$, and the average
number of operations to recover the $k$ source symbols is $O(k
\log(1/\epsilon))$.
\end{lemma}

\section{Raptor Codes Based Distributed Storage (RCDS) Algorithms}\label{sec:RCDSalgs}

As shown in~\cite{lin07a,aly08a}, distributed LT codes are relatively simple
to implement.  Raptor codes take the
advantage of LT codes to decode a major fraction of $k$ source
packets within linear complexity, and use another error correcting
code to decode the remaining minor fraction also within linear
complexity by concatenating such an error correcting code and LT
code together~\cite{shokrallahi06}.

Nevertheless, it is not trivial to achieve this encoding mechanism
in a distributed manner. In this section, we propose two algorithms
for distributed storage based on Raptor codes. The first is called
RCDS-I, in which each node has knowledge of limited global
information. The second is called RCDS-II, which is a fully
distributed algorithm and does not require any global information.

\subsection{With Limited Global Information---RCDS-I}

In RCDS-I, we assume that each node in the network knows the value
of $k$---the number of sources, and the value of $n$---the number of
nodes. We use simple random walk~\cite{Ro95} for each source to
disseminate its information to the whole network. At each round,
each node $u$ that has packets to transmit chooses one node $v$
among its neighbors uniformly independently at random, and sends the
packet to the node $v$. In order to avoid local-cluster
effect---each source packet is trapped most likely by its neighbor
nodes--- at each node, we make acceptance of any a source packet equiprobable. To
achieve this, we also need each source packet to visit each node in
the network at least once.

\begin{definition}(Cover Time)
Given a graph $G$, let $T_{cover}(u)$ be the expected length of a
random walk that starts at node $u$ and visits every node in $G$ at
least once. The \emph{cover time} of $G$ is defined by
$T_{cover}(G)=\max_{u\in G}T_{cover}(u)$~\cite{Ro95}.
\end{definition}

\begin{lemma}[{Avin and Ercal~\cite{avin05}}]\label{Lemma:Cover-Time}
Given a random geometric graph with $n$ nodes, if it is a connected
graph with high probability, then $T_{cover}(G)=\Theta(n\log n)$.
\end{lemma}

Therefore, we can set a counter for each source packet and increase
the counter by one after each forward transmission until the counter
reaches some threshold $C_1n\log n$ to guarantee that the source
packet visits each node in the network at least once.

To perform the LDPC pre-coding mechanism for $k$ sources in a
distributed manner, we again use simple random walks to disseminate
the source packets. Each source node generates $b$ copies of its own
source packet, where $b$ follows distribution for randomized LDPC
codes $\Omega_{L}(d)$. After these $b$ copies are sent out  and
distributed uniformly in the network, each node among $m$ nodes
chosen as pre-coding output nodes absorbs one copy of this source
packet with some probability. In this way, we have $m$ pre-coding
output nodes, each of which contains a combined version of a random
number of source packets. Then, the above method can be applied for
these $m$ pre-coding output nodes as new sources to do distributed
Raptor encoding. In this way, we can achieve distributed storage
packets based on Raptor codes. The RCDS-I algorithm is described in
the following steps.


\begin{compactenum}
\item[(i)] Initialization Phase:
\begin{compactenum}
\item[(1)] Each node $u$ in the network draws a random number $d_c(u)$ according to the
distribution $\Omega_{r}(d)$ given
by~\eqref{eq:Raptor-distribution}.

\item[(2)] Each source node $s_i, i=1,\dots,k$ draws a random number $b(s_i)$ according to
the distribution of $\Omega_{L}(d)$ and generates $b(s_i)$ copies of
its source packet $x_{s_i}$ with its ID and a counter $c(x_{s_i})$
with initial value zero in the packet header and sends each of them
to one of $s_i$'s neighbors chosen uniformly at random.
\end{compactenum}

\item[(ii)] Pre-coding Phase:
\begin{compactenum}
\item[(1)] Each node of the remaining $n-k$ non-source nodes chooses to serve as a
redundant node with probability $\frac{m-k}{n-k}$. We call these
redundant nodes and the original source nodes as pre-coding output
nodes. Each pre-coding output node $w_j$ generates a random number
$a(w_j)$ according to distribution $\Omega_{c}(d)$ given by
$\Omega_{c}(d)=\Pr(a(w)=d)=\binom{k}{d}\left(\frac{E[b]}{m}\right)^d
\left( 1-\frac{E[b]}{m}\right)^{k-d}$, where $E[b]=\sum_b
b\Omega_{L}(b)$.

\item[(2)] Each node that has packets in its forward queue before the current round sends the
head of line  packet to one of its neighbors chosen uniformly at
random.

\item[(3)] When a node $u$ receives a packet $x$ with counter $c(x)<C_1n\log(n)$ ($C_1$
is a system parameter), the node $u$ puts the packet into its
forward queue and update the counter as $c(x)=c(x)+1$.

\item[(4)] Each pre-coding output node $w$ accepts the first $a(w)$ copies of different
$a(w)$ source packet with counters $c(x)\geq C_1n\log(n)$, and
updates $w$'s pre-coding result each time as $y_w^+ = y_w^- \oplus
x$. If a copy of $x_{s_j}$ is accepted, the copy will not be
forwarded any more, and $w$ will not accept any other copy of
$x_{s_j}$. When the node $w$ finishes $a(w)$ updates, $y_w$ is the
pre-coding output of $w$
\end{compactenum}

\item[(iii)] Raptor-coding Phase:
\begin{compactenum}
\item[(1)] Each pre-coding output node $o_j$ put its ID and a counter $c(y_{o_j})$ with
initial value zero in the packet header, and sends out its
pre-coding output packet $y_{o_j}$ to one of its neighbor $u$,
chosen uniformly at random among all its neighbors
$\mathcal{N}(o_j)$.

\item[(2)] The node $u$ accepts this pre-coding output packet $y_{o_j}$ with probability
$\frac{d_c(u)}{m}$ and updates its storage as $z_u^+ = z_u^- \oplus
y_{o_j}$. No matter the source packet is accepted or not, the node
$u$ puts it into its forward queue
    and set the counter of $y_{o_j}$ as $c(y_{o_j})=1$.

\item[(3)] In each round, when a node $u$ has at least one pre-coding output packet in its forward
queue before the current round, $u$ forwards the head of line packet
$y$ in its forward queue to one of its neighbor $v$, chosen
uniformly at random among all its neighbors $\mathcal{N}(u)$.

\item[(4)] Depending on how many times $y$ has visited $v$, the node $v$ makes its decisions:
\begin{compactenum}
\item[\textbullet] If it is the first time that $y$ visits $u$, then the node $v$ accepts
this source packet with probability $\frac{d_c(v)}{m}$ and updates
its storage as $z_v^+ = z_v^- \oplus y$.

\item[\textbullet] If $y$ has visited $v$ before and $c(y)< C_1n\log n$, then the node
$v$ accepts this source packet with probability 0.

\item[\textbullet] No matter $y$ is accepted or not, the node $v$ puts it into its forward queue
    and increases the counter of $y$ by one $c(y)=c(y)+1$.

\item[\textbullet] If $y$ has visited $v$ before and $c(y)\geq C_1n\log n$ then the node
$v$ discards packet $y$ forever.
\end{compactenum}

\end{compactenum}

\item[(iv)] Storage Phase: When a node $u$ has made its decisions for all the pre-coding
output packets $y_{o_1},y_{o_1},...,y_{o_m}$, i.e., all these
packets have visited the node $u$ at least once, the node $u$
finishes its encoding process and $z_u$ is the storage packet of
$u$.
\end{compactenum}

The RCDS-I algorithm achieves the same decoding performance as
Raptor codes. Due to the space limitation, all the proofs for the
theorems and lemmas are omitted.

\begin{theorem}\label{Theorem:Decoding-RCDS-I}
Suppose sensor networks have $n$ nodes and $k$ sources, and let $k/m
= (1+\epsilon/2)/(1 +\epsilon)$. When $n$ and $k$ are sufficient
large, the $k$ original source packets can be recovered from
$(1+\epsilon)k$ storage packets. The decoding complexity is $O(k
\log(1/\epsilon))$.
\end{theorem}

The price for the benefits we achieved in the RCDS-I algorithm is
the extra transmissions. The total number of transmissions (the
total number of steps of $k$ random walks) is given in the following
theorem.

\begin{theorem}\label{Theorem:Transmission-RCDS-I}
Denote by $T_{RCDS}^{(I)}$ the total number of transmissions of the
RCDS-I algorithm, then we have
\begin{equation}\label{eq:T-RCDS-I}
T_{RCDS}^{(I)}=\Theta(kn\log n)+\Theta(mn\log n),
\end{equation}
where $k$ is the total number of sources before pre-coding, $m$ is
the total number of outputs after pre-coding, and $n$ is the total
number of nodes in the network.
\end{theorem}

\subsection{With no Global Information---RCDS--II}

In RCDS-I algorithm, we assume that each node in the network knows
$n$ and $k$---the total number of nodes and sources. However, in
many scenarios, especially, when changes of network topologies may
occur due to node mobility or node failures, the exact value of $n$
may not be available for all nodes. On the other hand, the number of
sources $k$ usually depends on the environment measurements, or some
events, and thus the exact value of $k$ may not be known by each
node either. As a result, to design a fully distributed storage
algorithm which does not require any global information is very
important and useful. In this subsection, we propose such an
algorithm based on Raptor codes, called RCDS-II. The idea behind
this algorithm is to utilize some features of simple random walks to
do inference to obtain individual estimations of $n$ and $k$ for
each node.

To begin, we introduce the definition of inter-visit time and
inter-packet time. For a random walk on any graph, the
\emph{inter-visit time} is defined as follows~\cite{Ro95, MoRa95}:
\begin{definition}(Inter-Visit Time)
For a random walk on a graph, the \emph{inter-visit time} of node
$u$, $T_{visit}(u)$, is the amount of time between any two
consecutive visits of the random walk to node $u$. This inter-visit
time is also called \emph{return time}.
\end{definition}

For a simple random walk on random geometric graphs, the following
lemma provides results on the expected inter-visit time of any node.
\begin{lemma}\label{Lemma:Inter-Visit-Time}
For a node $u$ with node degree $d_n(u)$ in a random geometric
graph, the mean inter-visit return time is given by
\begin{equation}\label{eq:E-T-visit-u}
E[T_{visit}(u)]=\frac{\mu n}{d_n(u)},
\end{equation}
where $\mu$ is the mean degree of the graph.
\end{lemma}

From Lemma~\ref{Lemma:Inter-Visit-Time}, we can see that if each
node $u$ can measure the expected inter-visit time
$E[T_{visit}(u)]$, then the total number of nodes $n$ can be
estimated by
\begin{equation}
\hat{n}'(u) = \frac{d_n(u)E[T_{visit}(u)]}{\mu}.
\end{equation}
However, the mean degree $\mu$ is a global information and may be
hard to obtain. Thus, we make a further approximation and let the
estimation of $n$ by the node $u$ be
\begin{equation}
\hat{n}(u) = E[T_{visit}(u)].
\end{equation}

In our distributed storage algorithms, each source packet follows a
simple random walk. Since there are $k$ sources, we have $k$
individual simple random walks in the network. For a particular
random walk, the behavior of the return time is characterized by
Lemma~\ref{Lemma:Inter-Visit-Time}. Nevertheless,
Lemma~\ref{Lemma:Inter-Packet-Time} provides results on the
inter-visit time among all $k$ random walks, which is called
inter-packet time for our algorithm and defined as follows:
\begin{definition}(Inter-Packet Time)
For $k$ random walks on a graph, the \emph{inter-packet time} of
node $u$, $T_{packet}(u)$, is the amount of time between any two
consecutive visits of those $k$ random walks to node $u$.
\end{definition}

\begin{lemma}\label{Lemma:Inter-Packet-Time}
For a node $u$ with node degree $d_n(u)$ in a random geometric graph
with $k$ simple random walks, the mean inter-packet time is given by
\begin{equation}\label{eq:E-T-packet-u}
E[T_{packet}(u)]=\frac{E[T_{visit}(u)]}{k}=\frac{\mu n}{kd_n(u)},
\end{equation}
where $\mu$ is the mean degree of the graph.
\end{lemma}

From Lemma~\ref{Lemma:Inter-Visit-Time} and
Lemma~\ref{Lemma:Inter-Packet-Time}, it is easy to see that for any
node $u$, an estimation of $k$ can be obtained by
\begin{equation}
\hat{k}(u)=\frac{E[T_{visit}(u)]}{E[T_{packet}(u)]}.
\end{equation}

After obtaining estimations for both $n$ and $k$, we can employ
similar techniques used in RCDS-I to perform Raptor coding and
storage. We will only present details of the Interference Phase due
to the space limitation. The Initialization Phase, Pre-coding Phase,
Raptor-coding Phase and Storage Phase are the same as in RCDS-I with
replacements of $k$ by $\hat{k}(u)$ and $n$ by $\hat{n}(u)$
everywhere.

Inference Phase:
\begin{compactenum}
\item[(1)] For each node $u$, suppose $x_{s(u)_1}$ is the first source packet that visits
$u$, and denote by $t_{s(u)_1}^{(j)}$ the time when $x_{s(u)_1}$ has
its $j$-th visit to the node $u$. Meanwhile, each node $u$ also
maintains a record of visiting time for each other source packet
$x_{s(u)_i}$ that visited it. Let $t_{s(u)_i}^{(j)}$ be the time
when source packet $x_{s(u)_i}$ has its $j$-th visit to the node
$u$. After $x_{s(u)_1}$ visiting the node $u$ $C_2$ times, where
$C_2$ is system parameter which is a positive constant, the node $u$
stops this monitoring and recoding procedure. Denote by $k(u)$ the
number of source packets that have visited at least once upon that
time.

\item[(2)] For each node $u$, let $J(s(u)_i)$ be the number of visits of source packet
$x_{s(u)_i}$ to the node $u$ and let $T_{s(u)_i}=
\frac{1}{J(s(u)_i)}
\sum_{j=1}^{J(s(u)_i)}t_{s(u)_i}^{(j+1)}-t_{s(u)_i}^{(j)}$. Let
$J_{ii'}\triangleq\min\{J(s(u)_i),J(s(u)_{i'})\}$, and
$T_{s(u)_is(u)_{i'}}= \frac{1}{J_{ii'}}
\sum_{j=1}^{J_{ii'}}t_{s(u)_i}^{(j)}-t_{s(u)_{i'}}^{(j)}$. Then, the
average inter-visit time and inter-packet time for node $u$ are
given by $\bar{T}_{visit}(u)= \frac{1}{k(u)}
\sum_{i=1}^{k(u)}T_{s(u)_i}$, and $\bar{T}_{packet}(u)=
\frac{1}{k(u)(k(u)-1)}
\sum_{i=1}^{k(u)-1}\sum_{i'=i+1}^{k(u)}T_{s(u)_is(u)_{i'}}$,respectively.
Then the node $u$ can estimate the total number of nodes in the
network and the total number of sources as
$\hat{n}(u)=\frac{\bar{T}_{visit}(u)}{2}$,and
$\hat{k}(u)=\frac{\bar{T}_{visit}(u)}{\bar{T}_{packet}(u)}$.

\item[(3)] In this phase, the counter $c(x_{s_i})$ of each source packet $c(x_{s_i})$ is
incremented by one after each transmission.
\end{compactenum}

\section{Performance Evaluation}\label{sec:performance}

In this section, we study the performance of the proposed  RCDS-I
and RCDS-II algorithms for distributed storage in wireless sensor
networks through simulation. The main performance metric we
investigate is the successful decoding probability versus the
decoding ratio.

\begin{definition}(Decoding Ratio)
\emph{Decoding ratio} $\eta$ is the ratio between the number of
querying nodes $h$ and the number of sources $k$, i.e.,
$\eta=\frac{h}{k}$.
\end{definition}

\begin{definition}(Successful Decoding Probability)
\emph{Successful decoding probability} $P_s$ is the probability that
the $k$ source packets are all recovered from the $h$ querying
nodes.
\end{definition}

In our simulation, $P_s$ is evaluated as follows. Suppose the
network has $n$ nodes and $k$ sources, and we query $h$ nodes. There
are $\binom{n}{h}$ ways to choose such $h$ nodes, and we choose
$M=\frac{1}{10}\binom{n}{h}=\frac{n!}{10\cdot h!(n-h)!}$ uniformly
randomly samples of the choices of query nodes. Let $M_s$ be the
number of samples of the choices of query nodes from which the $k$
source packets can be recovered. Then, the successful decoding
probability is evaluated as $P_s=\frac{M_s}{M}$.

\begin{figure}[t!]
\centerline{ \subfigure[]{
\includegraphics[scale=0.25]{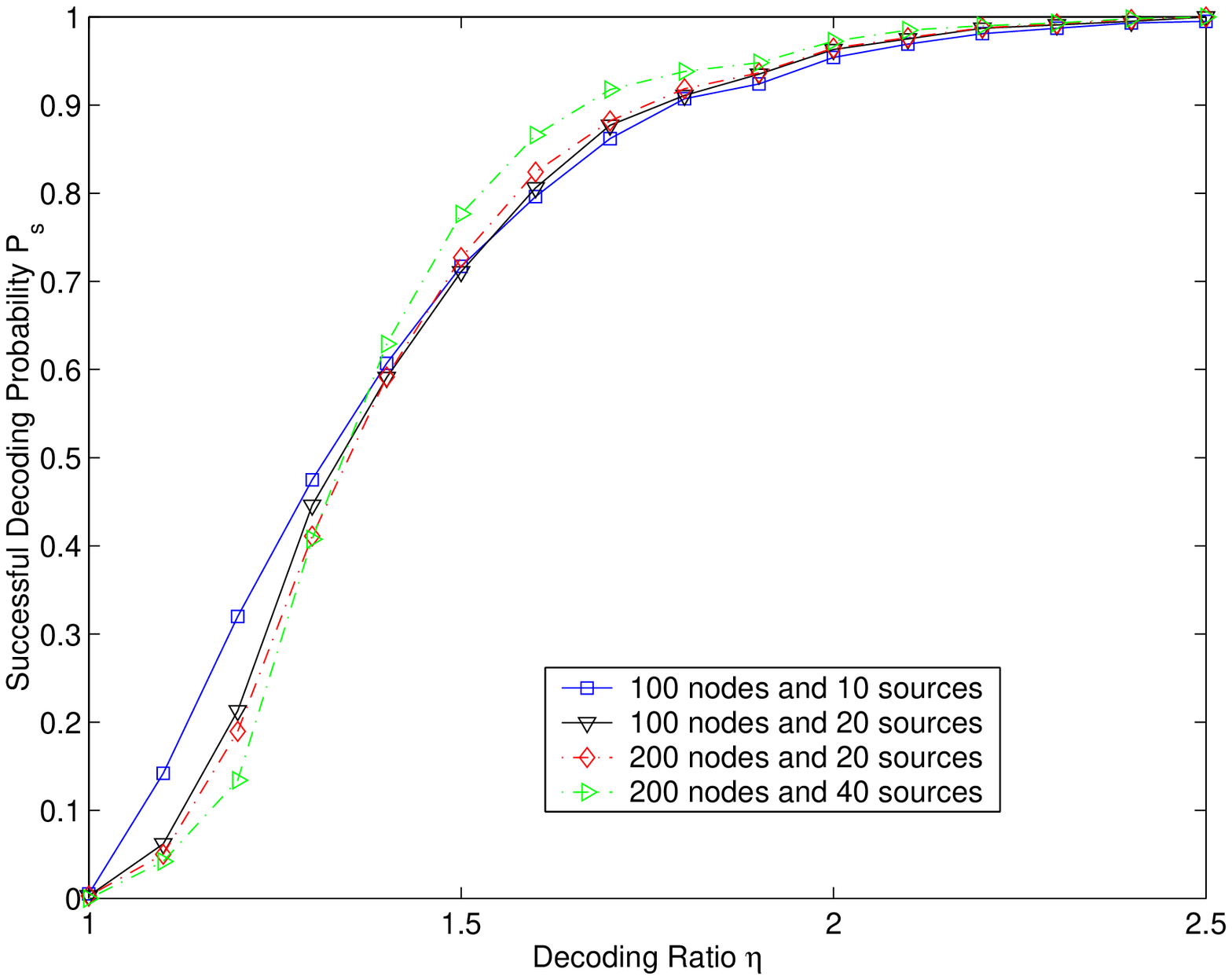}}\hfil
\subfigure[]{
\includegraphics[scale=0.25]{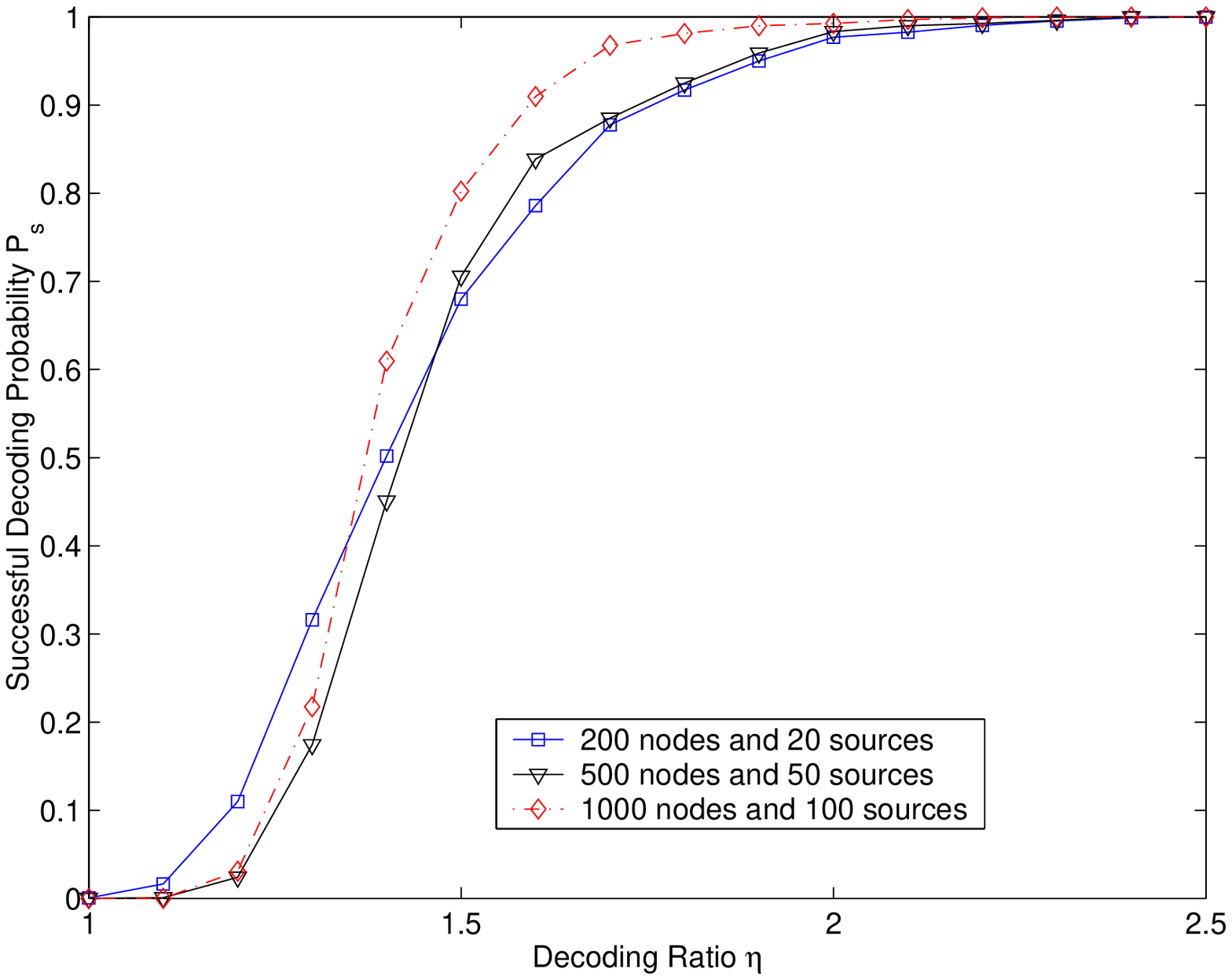}}}
\vspace{-.1in} \caption{Decoding performance of the RCDS-I
algorithm: (a) small number of nodes and sources; (b) large number
of nodes and sources}\label{fig:RCDS-I-1-2}\vspace{-.1in}
\end{figure}

\begin{figure}[t!]
\centerline{ \subfigure[]{
\includegraphics[scale=0.25]{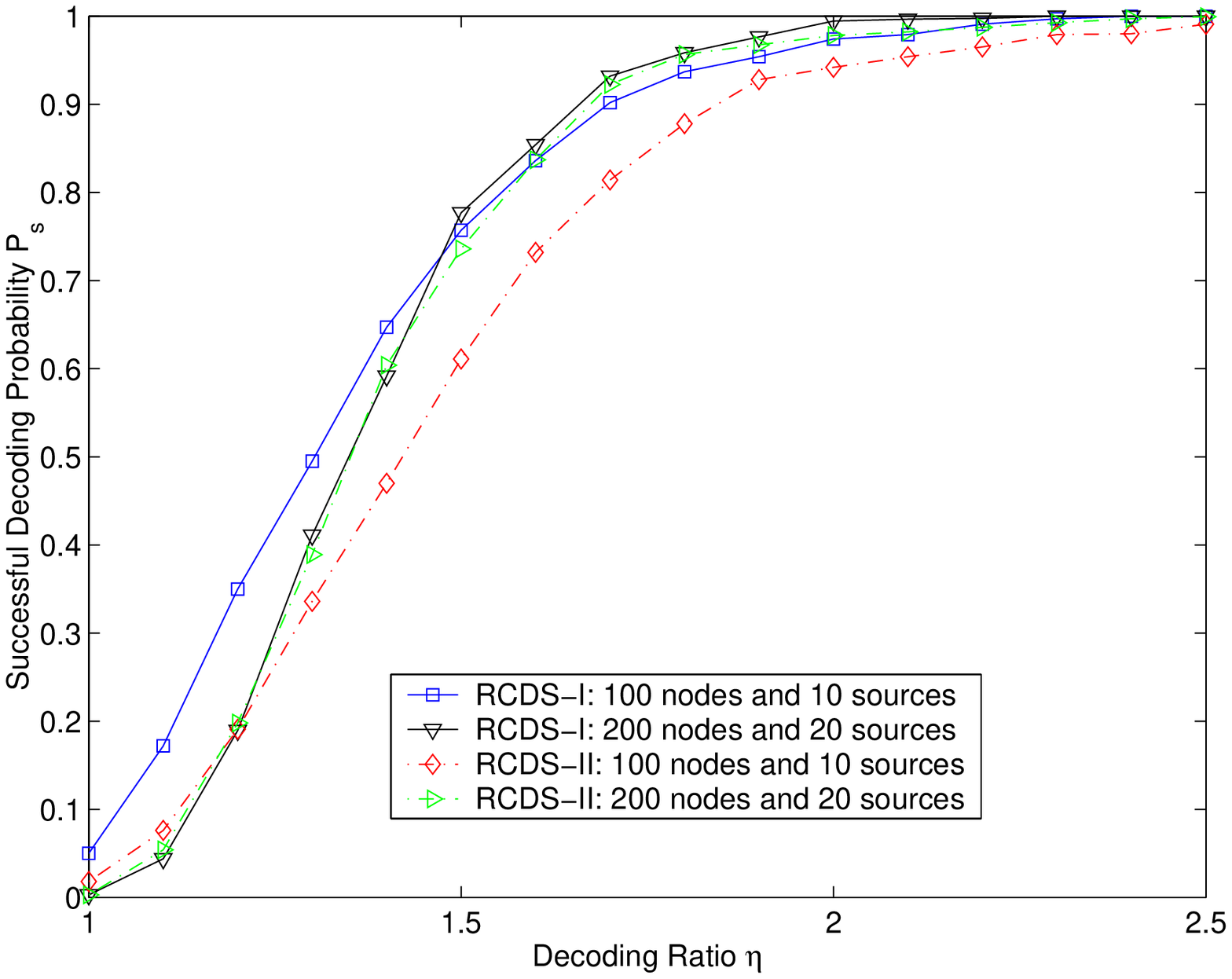}}\hfil
\subfigure[]{
\includegraphics[scale=0.25]{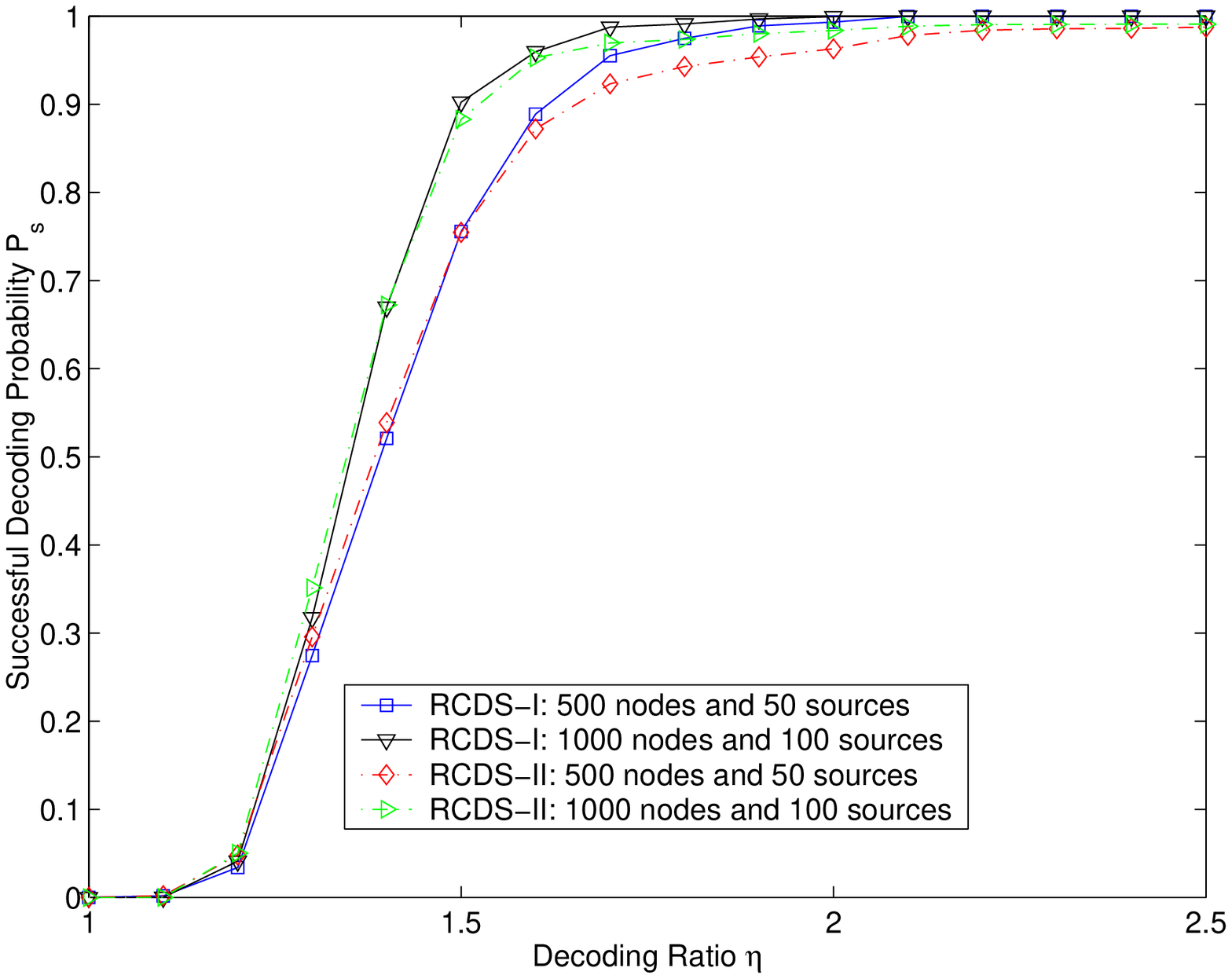}}}
\vspace{-.1in} \caption{Decoding performance comparison of the
RCDS-I and RCDS-II algorithms: (a) small number of nodes and
sources; (b) large number of nodes and
sources}\label{fig:RCDS-II-1-2}\vspace{-.1in}
\end{figure}

\begin{figure}[t!]
\centerline{ \subfigure[]{
\includegraphics[scale=0.25]{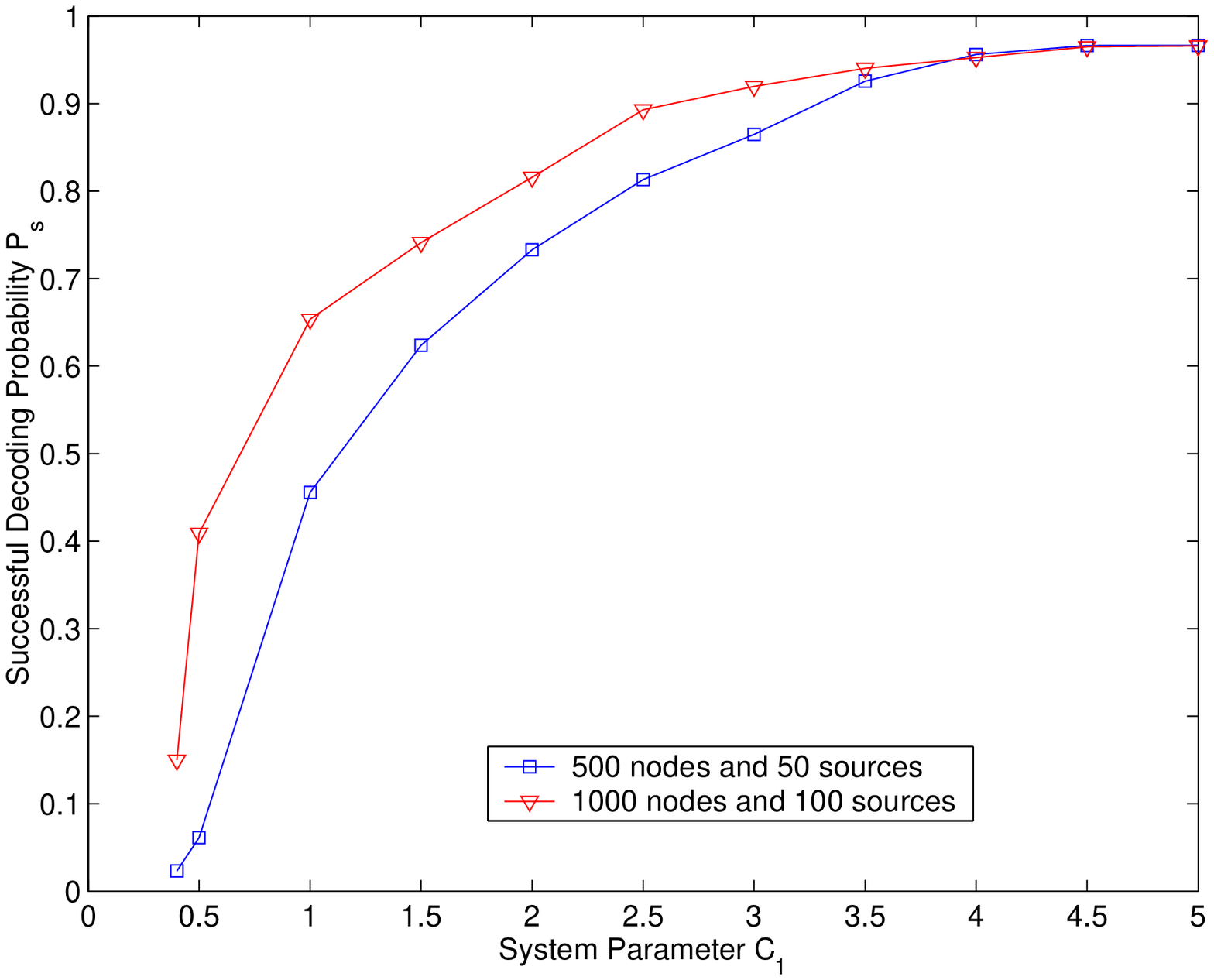}}\hfil
\subfigure[]{
\includegraphics[scale=0.25]{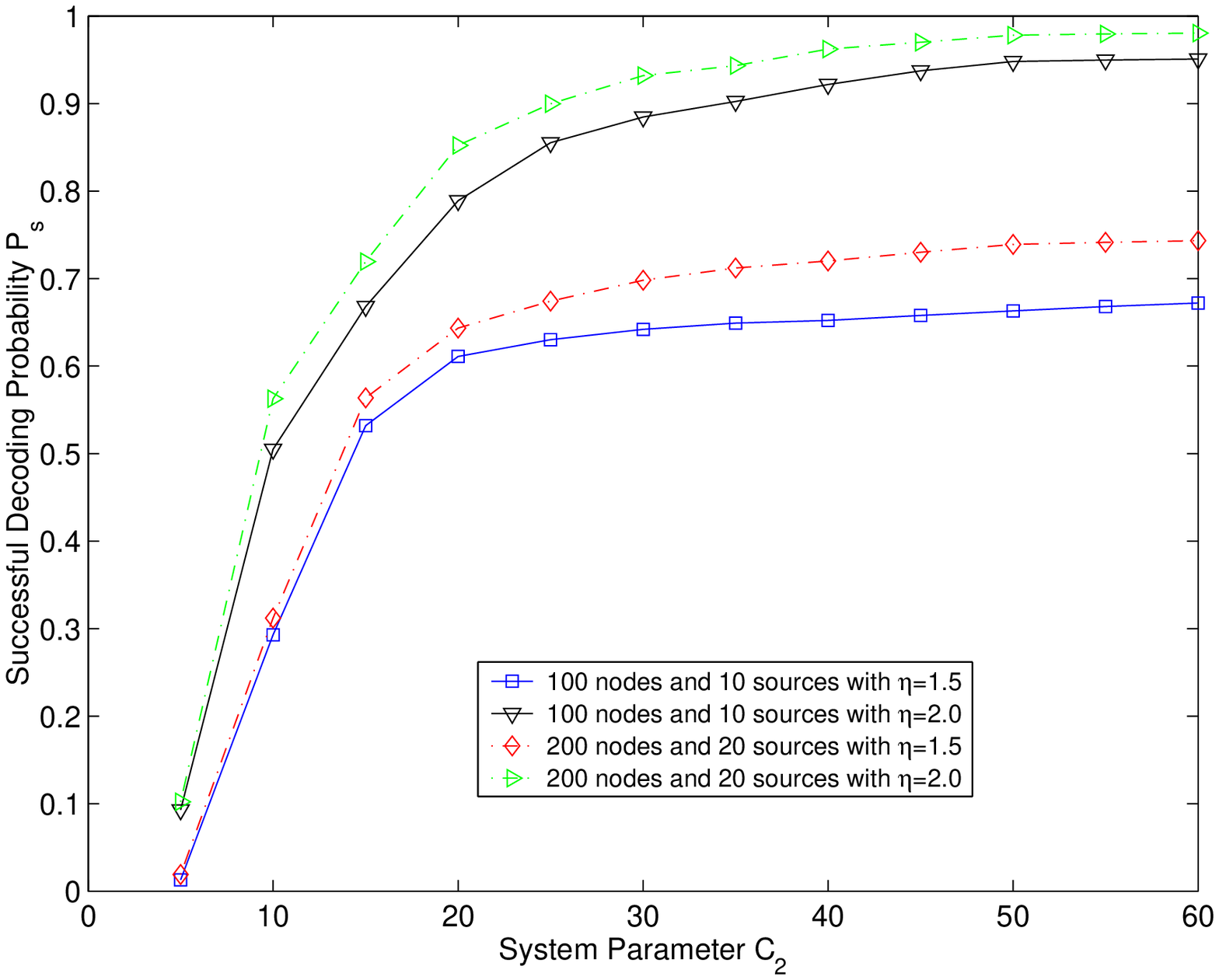}}}
\vspace{-.1in} \caption{Impact of system parameters: (a) decoding
performance of RCDS-I algorithm with different $C_1$, (b) decoding
performance of RCDS-I algorithm with different
$C_2$.}\label{fig:RCDS-I-II-3}\vspace{-.2in}
\end{figure}

Our simulation results are shown in Figures.~\ref{fig:RCDS-I-1-2},
~\ref{fig:RCDS-II-1-2} and~\ref{fig:RCDS-I-II-3}.
Fig.~\ref{fig:RCDS-I-1-2} shows the decoding performance of RCDS-I
algorithm with different number of nodes and sources. The network is
deployed in $\mathcal{A}=[5,5]^2$, and the system parameter $C_1$ is
set as $C_1=5$. From the simulation results we can see that when the
decoding ratio is above 2, the successful decoding probability is
about $95\%$. Another observation is that when the total number of
nodes increases but the ratio between $k$ and $n$ and the decoding
ratio $\eta$ are kept as constants, the successful decoding
probability $P_s$ increase when $\eta\geq 1.4$ and decreases when
$\eta<1.4$. That is because the more nodes we have, the more likely
each node has the desired degree distribution.
Fig.~\ref{fig:RCDS-II-1-2} compares the decoding performance of
RCDS-II and RCDS-I algorithms. To guarantee each node obtain
accurate estimations of $n$ and $k$, we set $C_2=50$. It can be seen
that the decoding performance of the RCDS-II algorithm is a little
bit worse than the RCDS-I algorithm when decoding ratio $\eta$ is
small, and almost the same when $\eta$ is large. To investigate how
the system parameter $C_1$ and $C_2$ affects the decoding
performance of the RCDS-I and RCDS-II algorithms, we fix the
decoding ratio $\eta$ and vary $C_1$ and $C_2$. The simulation
results are shown in Fig.~\ref{fig:RCDS-I-II-3}. It can be seen that
when $C_1\geq 4$, $P_s$ keeps almost like a constant, which
indicates that after $4n\log n$ steps, almost all source packet
visit each node at least once. We can also see that when $C_2$ is
chosen to be small, the performance of the RCDS-II algorithm is very
poor. This is due to the inaccurate estimations of $k$ and $n$ of
each node. When $C_2$ is large, for example, when $C_2\geq 40$, the
performance is almost the same.

\section{Conclusion}

In this paper, we studied Raptor codes based distributed storage
algorithms for large-scale wireless sensor networks. We proposed two
new decentralized algorithms RCDS-I and RCDS-II that distribute
information sensed by $k$ source nodes to $n$ nodes for storage
based on Raptor codes. In RCDS-I, each node has limited
global information; while in RCDS-II, no global information is
required. We computed the computational encoding and decoding
complexity, and transmission costs of these algorithms. We also
evaluated their performance by simulation.



\end{document}